# Template-free precision synthesis of biphilic microdome arrays

Haobo Xu[1], Haonian Shu[2] and Rong Yang[2,*]

[1]Department of Material Science and Engineering, Cornell University, Ithaca, New York, 14853, United States

[2]Robert Frederick Smith School of Chemical and Biomolecular Engineering, Cornell University, Ithaca, New York, 14853, United States

*Corresponding author: Rong Yang
**Email:** ryang@cornell.edu

**Abstract**
Biphilic microdome arrays are ubiquitous in nature, but their synthetic counterparts have been scarce. To bridge that gap, we leverage condensed droplet polymerization (CDP) to enable their template-free synthesis. During CDP, monomer droplets serve as microreactors for free-radical polymerization. The droplet diameter is monitored in real time. To enable the precise prediction of the convex geometric parameters, we develop a theoretical framework that integrates geometric arguments, scaling analysis, and kinetic theories. The model accurately predicts the dimensions of the as-synthesized microdome arrays, pointing to unprecedented precision in the synthesis of such topography. To illustrate its impact, the methodology is used to enable biphilic microdomes with targeted dimensions, for the reduction of surface colonization by a biofilm-forming pathogen, *Pseudomonas aeruginosa*. Importantly, the reduction is achieved with a moderately hydrophobic surface that has been considered prone to fouling, pointing to a fresh material design strategy and broadened palette of synthetic surfaces.

**Introduction**
Surfaces presenting microdome morphologies are ubiquitous in nature, with examples spanning all biological kingdoms, from bacterial biofilms(*1*) to plant seeds(*2*), crustacean exoskeletons, and mammal epithelial tissues. While the structure-function relationship in each instance is not yet fully understood, surfaces that exhibit a combination of convex features and contrasting surface energy, namely biphilic microdome surfaces, have demonstrated distinct functional benefits. For example, Namib desert beetles leverage these features for efficient water collection(*3, 4*), while spider silk presents spindle-knots and interspersed joints for directed liquid transfer(*5*). Nevertheless, it has been challenging to replicate such complex natural biphilic morphologies in synthetic materials.

This synthesis challenge largely arises from the need for precise dome geometries to achieve targeted functions. In Namib desert beetles, for example, the diffusive flux of water vapour toward the apex of domes scales with the radius of curvature during water condensation and collection(*6*). Similarly, the engineering of the Laplace pressure along spider silk hinges on its precise dome size and curvature. Despite substantial advancements in micro/nanofabrication and additive manufacturing, creating synthetic biphilic microdome arrays with targeted geometries remains a challenge. Techniques such as lithography(*7-9*), inkjet printing(*10, 11*), and electrospraying(*12*), while enabling a range of biphilic surfaces(*12-21*), have had limited success in producing convex surfaces with the targeted size, curvature, and reproducibility(*22*).

To enable the precision synthesis of biphilic microdome arrays, we employed an emerging technique called condensed droplet polymerization (CDP)(*23, 24*). The CDP process unfolds in three steps: (1) dropwise condensation of monomer vapor onto a hydrophobic substrate, (2) free radical polymerization within the condensed monomer droplets, and (3) evaporation of unreacted precursors from the resulting polymer-monomer mixture droplets (Fig. 1A). While condensation and evaporation are not commonly regarded as precision processes, here we develop a physics-based framework that integrates scaling-based analytical modeling and kinetic theories of droplet evaporation, gas diffusion, and polymerization that mitigates those imprecisions. Experimental validation confirms that the model enables accurate predictions of the size and curvature of the polymeric microdomes by connecting the geometric parameters to the droplet base radius at the onset of evaporation ($r_{b,i}$), which is monitored and controlled in real time(*25*). As a result, the exact geometric attributes of the polymeric microdomes can be systematically predicted and tailored, offering a pathway to the one-step precision synthesis of biphilic convex microstructures.

Leveraging the controlled synthesis, we explored the application of the biphilic microdome arrays in reducing the surface colonization by *Pseudomonas aeruginosa*, an opportunistic nosocomial pathogen. The biphilic topography, comprising microdomes of poly(2-hydroxyethyl methacrylate) (pHEMA) on a background of poly(*1H,1H,2H,2H*-perfluorodecyl acrylate) (pPFDA), led to minimal adhesion by *P. aeruginosa*. We emphasize that the biphilic microdome arrays are distinct from existing antifouling surfaces, which present either wetting contrast without 3D topography (e.g., amphiphilic copolymers)(*26-30*) or surface structures (e.g., sharklet) without wetting contrast(*31-33*). Furthermore, the pHEMA microdome arrays reduce biofouling while retaining an overall hydrophobic surface, as indicated by water contact angle measurements, which defies the long-held belief that treats hydrophilicity as the yardstick in antifouling material design. Beyond pHEMA, the synthesis method is compatible with a wide range of functional polymers, whose microdome arrays can be obtained simply by varying the monomer selection(*23*). This potential points to the facile synthesis of functional dome-shaped surfaces and particles with programmable geometric parameters, which have been shown to dictate their performance and functions in drug delivery(*34-41*), injectable implants(*42*), nano-optics(*43-45*), and soft robotics(*46*). The precision synthesis capability reported here meets the demand for simple one-step synthesis and precision geometric control, which is poised to accelerate the development of shape-engineered emergent materials.

## Results

**Selection of conditions and precursors to build the quantitative model for the precision synthesis of biphilic microdome arrays**

CDP is conducted by placing the substrate (pPFDA-coated silicon wafer) on a thermoelectric cooler (TEC) module kept in a cylindrical vacuum reactor(*23-25, 47*). The three steps of CDP, i.e., condensation, polymerization, and evaporation (Fig. 1A), are performed sequentially once the reactor base pressure is reached (P ≤ 5 mTorr). Despite the unique capability of CDP to produce nano-to-micro-sized convex surface structures, the dome arrays synthesized thus far have exhibited varying sizes and aspect ratios with limited precision(*23, 25*). To develop a quantitative model that enables the precise determination of the geometric parameters, we focus on the polymerization and evaporation step, which collectively determine the microdome morphology for a given monomer droplet. As such, to simplify the model construction, we leveraged the in-situ microscope to achieve monomer droplet sizes of 20~40 μm as the common starting point of the CDP procedures in this study. Furthermore, to capture the details of the morphological evolution during evaporation, we slowed down the evaporation of unreacted monomers using a two-step procedure: first, by applying vacuum, and then by increasing the TEC temperature. Using this adapted protocol, we successfully fabricated microdomes with a variety of sizes and geometries. The microdome morphology is represented quantitatively using the height-to-width ratio (Fig. 1B), which decreases with increasing dome diameter; e.g., for a dome with a diameter of around 10 μm, the height-to-width ratio is approximately 0.3, corresponding to a contact angle of around 60°. This trend will be analyzed in detail in the following sections.

Two monomers were used in this study, benzyl methacrylate (BzMA) and HEMA, and TBPO was used as the initiator for the free radical polymerization (Fig. 1C). The successful obtainment of pBzMA and pHEMA microdome arrays was confirmed using Fourier transform infrared (FTIR) spectroscopy, scanning electron microscopy (SEM), and energy-dispersive X-ray (EDX) spectroscopy. The SEM and EDX images illustrate that the pBzMA domes (dBzMA) are chemically distinct from the coating layer, as evidenced by the absence of fluorine where domes were observed (Fig. 1D). Additionally, the FTIR spectra of pPFDA coating layer, dBzMA atop of coating layer, and pBzMA thin films (Fig. 1E) show that peaks corresponding to both pPFDA (black) and pBzMA (blue) are preserved in the dBzMA surface (purple), with peaks unique to pBzMA highlighted in gray shade. The broad peak at 2955 $cm^{-1}$ identifies aromatic C–H stretch, and the peak at 695 $cm^{-1}$ identifies out-of-plane bending of the C–H bond in the benzene ring(*48*). The same set of characterizations was performed on the pHEMA domes (dHEMA). The absence of fluorine and strong signal of oxygen at the dome positions confirm that the domes are chemically distinct and reside atop the coating layer (Fig. 1F). In the FTIR spectra (Fig. 1G), the broad peak above 3000 $cm^{-1}$ identifies the O−H stretching in dHEMA/pHEMA, and the peak at 1457 $cm^{-1}$ corresponds to the bending of the C−H bond in the polymer backbone and the ethyl moiety of the dHEMA/pHEMA side chains(*49, 50*).

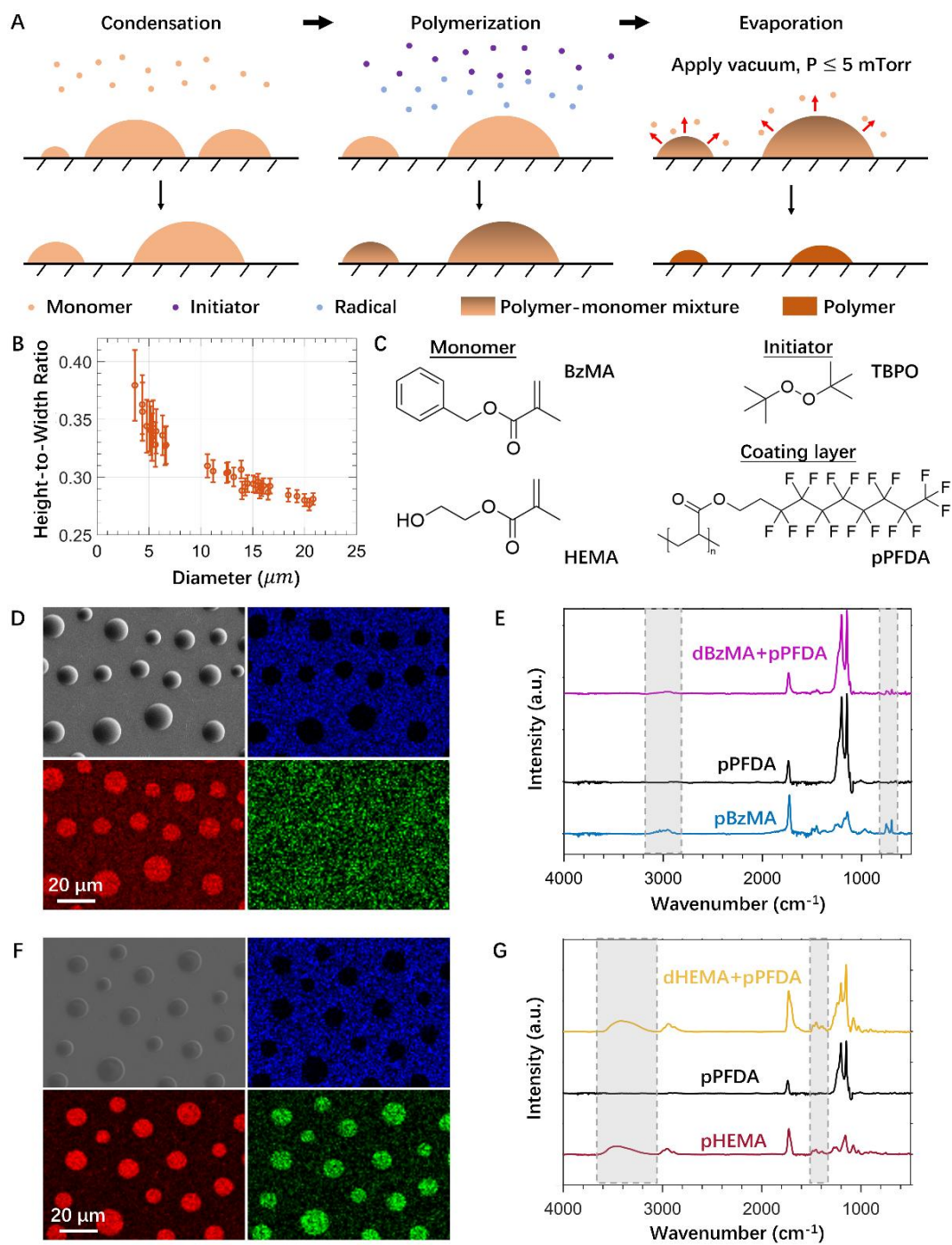

**Fig. 1 | CDP process, dome geometry, and chemical composition analysis.** (**A**) Three steps of CDP: monomer condensation, polymerization, and evaporation. (**B**) Plot of the dome height-to-width ratio versus diameter, with an inverse correlation. (**C**) Molecular structures of the monomers used to synthesize the microdome arrays (i.e., BzMA and HEMA), the polymerization initiator (TBPO), and the coating layer (i.e., pPFDA) applied to substrates as a base layer to promote dropwise monomer condensation. (**D**) SEM image of dBzMA (top left) with EDX mapping of fluorine (top right), carbon (bottom left), and oxygen (bottom right). (**E**) FTIR spectra of pBzMA (i.e., thin film, blue), pPFDA (i.e., the base layer, black), and dBzMA (i.e., pBzMA domes on the base layer, purple). The gray shading highlights the pBzMA peaks that are not present in the base layer. (**F**) SEM image of dHEMA (top left) with EDX mapping of fluorine (top right), carbon (bottom left), and oxygen (bottom right). (**G**) FTIR spectra of pHEMA (i.e., thin film, red), pPFDA (i.e., the base layer, black), and dHEMA (i.e., pHEMA domes on

the base layer, gold). The gray shading highlights the pHEMA peaks that are not present in the base layer.

**Three-stage evaporation during CDP and its role in dome shape formation**

As discussed above, we focus on the polymerization and evaporation steps to predict the geometric parameters of the domes formed by CDP. Polymerization determines the amount of polymer synthesized, whereas evaporation directly governs the dome formation process. Below, we first discuss the evaporation process as the mode of evaporation influences the droplet contact angle, which in turn affects the final dome shape.

A custom Python script was developed to track changes in the average droplet size during the evaporation process by detecting contours (e.g., the orange outline in Fig. 2A). Based on the temporal evolution of the average radius shown in Fig. 2B, we divide the process into three stages: (I) gradual shrinkage, (II) major evaporation, and (III) dome formation. The average radius corresponding to the four frames shown in Fig. 2A are marked by red circles in Fig. 2B. Stage I is driven by evacuating the reactor chamber to a base pressure of 5 mTorr (t = 0–40 s, stage I in Fig. 2B). During this stage, minimal evaporation occurs since the substrate remains cooled (~10°C), resulting in a low saturation vapor pressure of monomer ($P_{v,1}$ in Fig. 2C). The conditions corresponding to this stage is represented by point A in the phase diagram (Fig. 2C). As such, $\Delta P_1$, i.e., the pressure difference between the monomer's saturation pressure and the chamber pressure, is small ($\approx$ 4 mTorr), providing a limited driving force for evaporation (see detailed calculation in Supplementary Note 1). Stage II is driven by increasing the temperature of the substrate to ~20°C, with the corresponding conditions represented by point B in Fig. 2C. The difference between the chamber pressure and the monomer saturation pressure under this substrate temperature, $\Delta P_2 \approx$ 18 mTorr, thereby increasing the driving force of monomer evaporation. Consequently, in stage II (t = 40–90 s in Fig. 2B), the droplets shrink at a significantly accelerated rate and the majority of the unreacted monomer evaporates. Eventually, monomer evaporation completes and solid domes form, reaching a plateau in the average droplet radius (t > 110 s, stage III in Fig. 2B).

During stages I and II, the squared average droplet base radius ($r_b^2$) exhibits a linear relationship with time, suggesting that evaporation likely occurs in a constant contact angle (CCA) mode, which has been described using the relationship as follows in traditional droplet evaporation theory(*51*):

$$r_b^2 = r_{bi}^2 - \frac{4D(c_S - c_\infty)sin^2\theta}{\rho_L(1 - cos\theta)(2 + cos\theta)}t \tag{1}$$

where $r_{bi}$ is the initial droplet base radius, $D$ is the self-diffusion coefficient of the monomer vapor, $c_S$ is the monomer vapor concentration at the droplet-vapor boundary, $c_\infty$ is the bulk vapor concentration in the chamber, $\theta$ is the droplet contact angle, $\rho_L$ is the density of the monomer. Throughout the evaporation process, the vapor diffusion coefficient $D$ and liquid density $\rho_L$ can be approximated to be constant. However, the vapor concentration $c_S$ changes when transitioning from stage I to stage II due to variations in the monomer saturation pressure, and it remains stable within each stage. The linearity of the observed $r_b^2 \sim t$ correlation implies that $\theta$ is relatively stable during stages I and II, i.e., the droplet evaporation follows the CCA mode. In other words, the contact angle of the domes emerging at the end of stage II approximates the receding angle of the liquid droplet, irrespective of the droplet or dome size. Nevertheless, as shown in Fig. 1B, the dome contact angle varies from 60° to 75°, depending on the dome radius, which contradicts the CCA assumption. The main reason is that as the polymer volume fraction increases during the evaporation, the droplet becomes very viscous and no longer behaves like a liquid. As such, in stage III, the contact angle and contact radius decrease simultaneously, making it challenging to predict the dome contact angle using traditional droplet evaporation theory. Consequently, our focus below shifts to analyzing the post-evaporation geometric dome shape and integrating the polymerization kinetics during step 2 in CDP to predict the dome size and shape.

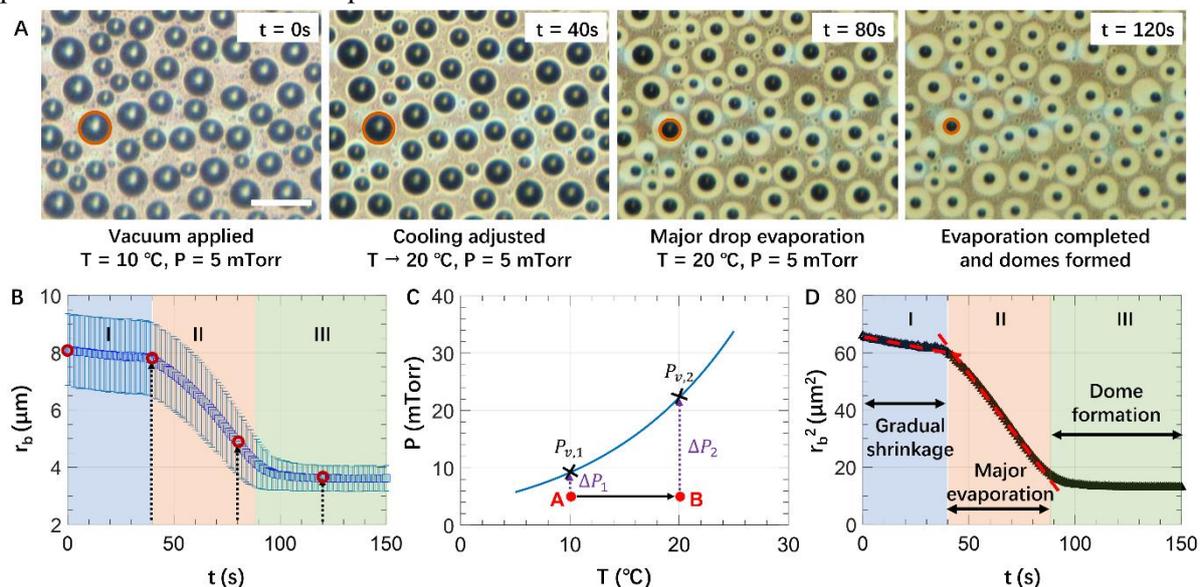

**Fig. 2 | Analysis of the evaporation step in CDP.** (**A**) Time-lapse images (scale bar = 50 μm) of droplets during the evaporation step captured at 15 fps. Orange circle shows the size tracking of a single droplet base radius during evaporation. Chamber evacuation begins at t = 0s, and the substrate temperature is increased to 20°C at t = 40s. (**B**) Plot of the average droplet base radius versus time. The red circle corresponds to the snapshots in (**B**). The error bar shows 95% confidence interval of the droplet base radius. (**C**) Vapor-liquid equilibrium plot for BzMA (blue line). The substrate temperature increases after the cooling adjustment, moving the state of the synthesis system from A to B, resulting in an increase in the saturation pressure of BzMA (from $P_{v,1}$ to $P_{v,2}$) and therefore an increase in the evaporation driving force, from $\Delta P_1$ to $\Delta P_2$. (**D**) Plot of squared average droplet base radius versus time. Droplet evaporation occurs in three stages: (I) gradual shrinkage, (II) major evaporation, and (III) dome formation. The red dotted lines in stages I and II indicate they are linear regions following the CCA mode.

**Geometric modeling and scaling law to predict microdome morphology based on direct shape measurement**

To fully account for the microdome/droplet morphology and its time-evolution, below we introduce several geometric parameters (Fig. 3A). At the onset of evaporation, the droplet (a mixture of monomer and polymer) has a base radius of $r_{bi}$ and a height of $H_i$; during the evaporation of unreacted monomer, the decreasing droplet base radius and height are represented by $r_b$ and $H$, respectively, and the final polymer dome base radius and height are represented by $r_{bf}$ and $H_f$.

Using the in-situ microscope, $r_{bi}$ can be observed and controlled in real time, and $r_{bf}$ can be quantified directly. While $H_i$ cannot be directly observed, the contact angle of monomers on pPFDA offers a close approximation, given the low conversion rates previous reported for CDP(24, 47). As $H_f$ is not directly observable and challeging to characterize in situ, to predict the size and shape of the as-synthesized microdome arrays, below we develop a theoretical framework that expresses $r_{bf}$ and $H_f$ as function of $r_{bi}$, the key observable and controllable quantity in CDP.

To analyze the final dome geometries accurately, we used an ex-situ Keyence VK-X260 Laser-Scanning Profilometer, which measures $r_{bi}$ (orange circle in Fig. 3B), $r_{bf}$ (purple circle in Fig. 3B) and the height profile $H$. The profilometer results were benchmarked against data obtained using atomic force microscopy (AFM), and the two methods showed excellent agreement (see Supplementary Fig. S1). As such, we leveraged the optical profilometer to obtain dome geometric data in a high-throughput fashion, which revealed that the dimensionless profile of the microdomes can be described by a unifying parabolic function. Below, we use two domes (labeled as I and II in Fig. 3B) to illustrate in detail how we arrived at that conclusion. The height profiles, i.e., $H$ vs $R$, for domes I and II are plotted in Fig. 3C. With the dome aspect ratio defined as $H_f/2r_{bf}$, dome I and II exhibit aspect ratios of 0.31 and 0.28 respectively, implying that larger droplets (dome II in this case) shrink more in both radial and vertical dimensions during evaporation. To verify this observation, we analyzed the geometric data of domes synthesized using different monomer species, monomer droplet size, and polymerization time, and confirmed that the radius ratio $r_{bf}/r_{bi}$ indeed decreases with $r_{bi}$ (Fig. 3D). By similarly nondimensionalizing the dome height profiles ($H/r_{bi}$ vs $R/r_{bi}$ in Fig. 3C), domes I and II nearly overlap, indicating that the dome profiles may be self-similar in nondimensional space. We assumed that the the relationship between dimensionless height $H_f/r_{bi}$ and dimensionless base radius $r_{bf}/r_{bi}$ follows a power law, and $H_f/r_{bi} = \lambda(r_{bf}/r_{bi})^2$ ($\lambda$ is a constant describing the parabolic dome profile) shows excellent agreement (Fig. 3E, Table S1) irrespective to the initial droplet size $r_{bi}$ (3~30 μm), monomer species (HEMA & BzMA) and polymerization time (1~4 min). From Fig. 3E, we obtain $\lambda$ = 0.75 directly from the slope, while from Supplementary Fig. S2 (data-fitting using equation (3), as detailed below), we can get $\lambda$ equals to 0.78. The agreement suggests that the dimensionless profile of domes are parabolic, which is commonly used in fitting evaporating droplet geometries(*52, 53*).

While the geometric relationship in dimensionless space provides a descriptive model of the dome profile, to establish a direct scaling law linking dome geometric factors ($r_{bf}$, $H_f$) with a controllable synthesis parameter ($r_{bi}$), below we leverage mass conservation of the polymer during evaporation. At the onset of evaporation, we denote the polymer volume fraction in a droplet (a mixture of polymer and monomer) as $\chi_i$. Thus, the mass of polymer is given by $\rho_{po}V_i\chi_i$, where $\rho_{po}$ is the polymer density, and $V_i$ is droplet volume before evaporation. At the end of evaporation, we assume all unreacted monomer is removed, leaving a polymer dome with a final volume of $\frac{\pi}{2}r_{bf}^2 H_f$. From mass conservation, we have:

$$\rho_{po}\frac{\pi}{2}r_{bf}^2 H_f = \rho_{po}V_i\chi_i = \rho_{po}\pi r_{bi}^3 \beta_i \chi_i \qquad (2)$$

where $\beta_i = (1-cos\theta_i)^2(2+cos\theta_i)/(3sin^3\theta_i)$, which is the geometric factor used to calculate the volume of a droplet, and $\theta_i$ is the contact angle of a monomer on pPFDA (86.1° for BzMA, 87.4° for HEMA, see Supplementary Fig. S3), giving rise to a relatively constant $\beta_i$ value of 0.59. By further simipifying equation (2) and combining the geometric relation $H_f/r_{bi} = \lambda(r_{bf}/r_{bi})^2$, we obtained the expression for dimensionless height and radius:

$$\frac{H_f}{r_{bi}} = (2*\beta_i*\lambda)^{\frac{1}{2}}\chi_i^{\frac{1}{2}} = (1.18*\lambda)^{\frac{1}{2}}\chi_i^{\frac{1}{2}} \qquad (3)$$

$$\frac{r_{bf}}{r_{bi}} = \left(2 * \frac{\beta_i}{\lambda}\right)^{\frac{1}{4}} \chi_i^{\frac{1}{4}} = \left(\frac{1.18}{\lambda}\right)^{\frac{1}{4}} \chi_i^{\frac{1}{4}} \tag{4}$$

Equation (3) and (4) give rise to the scaling laws for both dimensionless height $H_f/r_{bi} \sim \chi_i^{1/2}$ and dimensionless radius $r_{bf}/r_{bi} \sim \chi_i^{1/4}$, which are validated by experimental results (Fig. 3F,G).

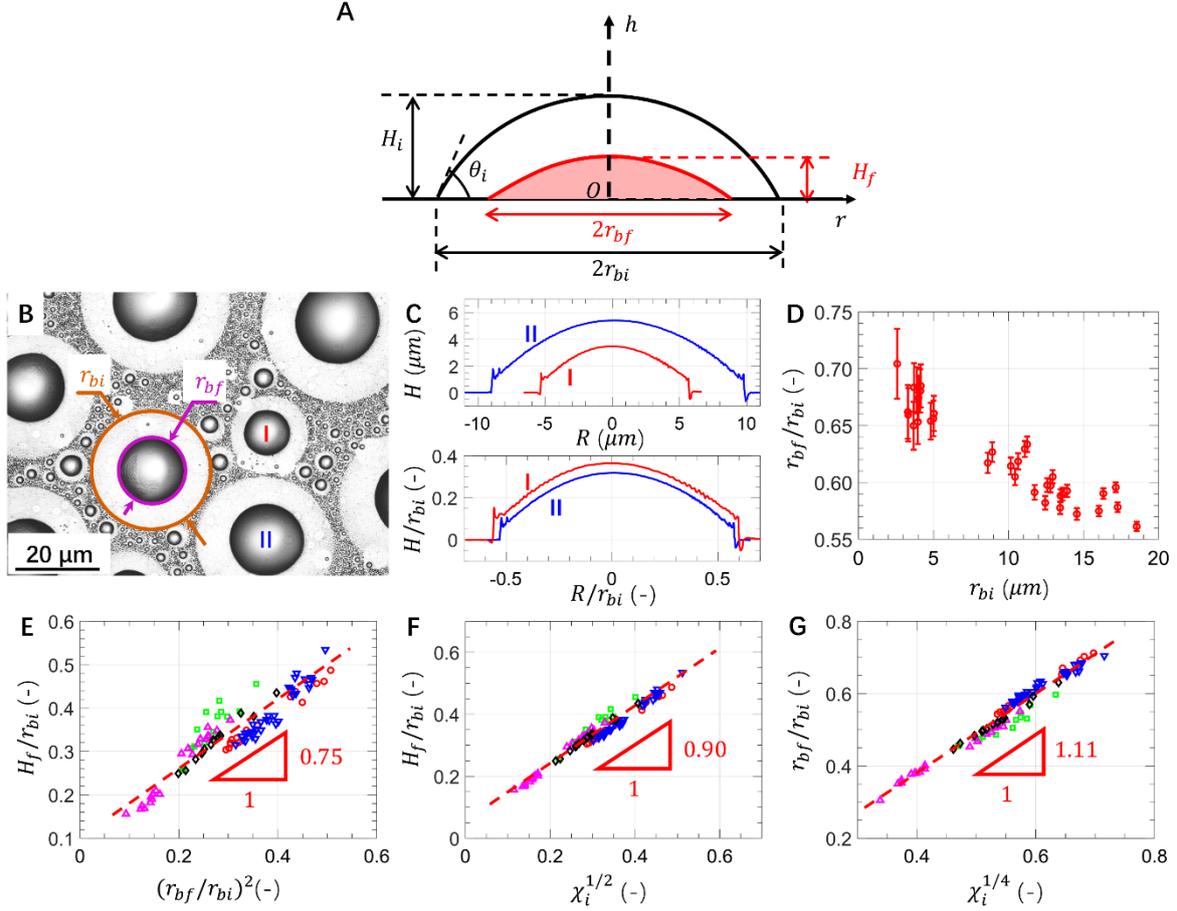

**Fig. 3 | Shape analysis for CDP domes.** (**A**) Schematic of the geometric parameters of a monomer droplet and the corresponding polymer dome formed during the CDP process. (**B**) Optical profilometer data to obtain the droplet base radius at the onset of evaporation ($r_{bi}$), the polymer dome base radius ($r_{bf}$) and polymer dome height ($H_f$). (**C**) Profile plots of the two domes highlighted in (**B**) (I in red; II in blue). The upper plot is height ($H$) versus radius ($R$), and the lower plot is dimensionless height ($H/r_{bi}$) versus dimensionless radius ($R/r_{bi}$). (**D**) Ratio between dome base radius $r_{bf}$ and the droplet base radius $r_{bi}$ versus droplet base radius $r_{bi}$, illustrating that larger $r_{bi}$ leads to a smaller $r_{bf}/r_{bi}$, i.e., larger droplets shrink more during evaporation. (**E**) Dimensionless dome height ($H_f/r_{bi}$) scales with the dimensionless dome base radius square ($r_{bf}/r_{bi}$)² with a slope of 0.75, irrespective of initial droplet radius, monomer species, and polymerization time, which are represented with different data symbols in the figure. (**F,G**) Dimensionless dome height ($H_f/r_{bi}$) and dimensionless dome base radius ($r_{bf}/r_{bi}$) scales with polymer volume fraction ($\chi_i$) to the power of 1/2 and 1/4, respectively, validating our prediction. See Table S1 for the synthesis conditions corresponding to each symbol.

**Modeling of the CDP kinetics for an analytical expression of the microdome geometry as a function of the initial droplet base radius $r_{bi}$**

From the geometric relationship and mass conservation, we obtain two scaling laws to relate the dome height and base raidus to the polymer volume fraction at the end of polymerization (step 2 in CDP). To predict the dome size and height accurately from droplet base radius ($r_{bi}$), a parameter monitored and controlled using the in-situ microscope, below we analyze the kinetics of the polymerization process to establish a relationship between the polymer volume fraction and $r_{bi}$.

We start from the generic expression for the rate of propagation during free radical polymerization:

$$R_p = -\frac{d[M]}{dt} = k_p[M \bullet][M] \quad (5)$$

where $R_p$ is the rate of propagation, $k_p$ is the propagation rate constant, $[M]$ is the monomer concentration and $[M \bullet]$ is the total concentration of all chain radicals. To eliminate $[M \bullet]$, we apply the pseudo-steady-state assumption which is applicable since CDP has a monomer conversion rate less than 30%(54). Therefore, by equating the rates of initiation $R_i$ and termination $R_t$, $R_i = R_t = 2k_t[M \bullet]^2$, where $k_t$ is the rate constant of termination, we obtain $[M \bullet] = (R_i/2k_t)^{1/2}$, and equation (5) becomes:

$$R_p = -\frac{d[M]}{dt} = k_p[M]\left(\frac{R_i}{2k_t}\right)^{\frac{1}{2}} \quad (6)$$

Equation (6) indicates a scaling law for rate of propagation and rate of initiation in monomer droplet.

During CDP, initiation comprises three steps that are spatially separated: (i) primary radical generation in the vapor phase, (ii) its diffusion and absorption to a monomer droplet, and (iii) its reaction with a monomer in the condensed phase. We note that this treatment is based on the recent evidence showing that the reaction of primary radical and monomer in the vapor phase is negligible(55).

In step (i), we estimate the rate of radical generation ($R_{iv}$) in the vapor phase (via thermolysis of the initiator, *tert*-butyl peroxide) using $R_{iv} = d[R \bullet]_v/dt = 2fk_d[I]$, where $[R \bullet]_v$ is the free radical concentration in vapor phase, $f$ is the initiator efficiency, $k_d$ is the rate constant of decomposition, $[I]$ is the concentration of initiator in the vapor phase.

To address steps (ii) and (iii), we adopt an empirical treatment commonly used in CVD polymerization literature, which treats the combined processes of absorption and reaction of the primary radical using sticking coefficient Γ. The sticking coefficient estimates the probability that a vapor-phase primary radical reacts with a monomer-covered surface(56, 57). As such, the overall rate of steps (ii) and (iii) can be estimated as the primary radical absorption rate times the sticking coefficient. The flux of primary radicals toward the surface of a droplet spherical cap depends on its mode of diffusion (Fig. 4A), and thus, we first calculate their mean free path under the experimental conditions, which is around 300 μm. The corresponding Knudsen number is 7.5, which is in a range where both collision-dominated (Knudsen diffusion) and diffusion-dominated (Fickian diffusion) models may be applicable.

Under Knudsen diffusion, assuming radicals are well-mixed in the vapor phase with isotropic velocity, the radical flux is given by $\Phi = [R \bullet]_v v N_A/4$, where $v$ is the average velocity of radicals. Therefore, for the number rate of primary radical absorption to a droplet spherical cap per unit time, we have $dN_R/dt = \Phi A = \Phi \pi r_{bi}^2 \alpha_i$, where $\alpha_i = 2(1 - \cos(\theta_i))/\sin^2\theta_i$ is a geometric factor for the spherical cap surface area calculation. The rate of initiation in a monomer droplet can thus be expressed as:

$$R_i = -\frac{d[R\bullet]}{dt} = \frac{1}{V_i}\frac{1}{N_A}\frac{dN_R}{dt}\Gamma = \frac{1}{r_b}[R\bullet]_v v \Gamma \frac{\alpha_i}{\beta_i} \tag{7}$$

where $\Gamma$ is the sticking coefficient(56) and $\beta_i$ is the geometric factor for droplet volume calculation (see equation (2)). By replacing $R_i$ in equation (6), rearrangement and integration gives rise to the unreacted monomer concentration after polymerization $[M]_{t_{po}}$:

$$[M]_{t_{po}} = [M]_0 e^{-\frac{k_C}{r_{bi}^{1/2}}} \tag{8}$$

where $t_{po}$ is the time of polymerization, $k_C$ is a composite kinetic parameter for the collision (Knudsen-diffusion) model that contains information on the kinetic constants discussed above (see details in Supplementary Note 5).

Under Fickian diffusion, the flux of primary radicals is given by $\Phi \sim D[R\bullet]_v/r_{bi}$, where $D$ is the diffusion coeffieicent of primary radicals in vapor phase. This results in a change in the scaling of $[M]_{t_{po}}$:

$$[M]_{t_{po}} = [M]_0 e^{-\frac{k_D}{r_{bi}}} \tag{9}$$

where $k_D$ is a composite kinetic parameter for the diffusion (Fickian-diffusion) model. The expressions for unreacted monomer concentration in equations (8) and (9) help derive polymer volume fraction ($\chi_i$, at the onset of evaporation) through a straightforward mass conservation calculation, where the mass of polymer plus the mass of unreacted monomer equals initial mass of the monomer droplet:

$$V_p \rho_{po} + V_i e^{-\frac{k}{r_{bi}^a}} \rho_{mo} = V_i \rho_{mo}$$

$$\chi_i = \frac{V_p}{V_i} = \frac{\rho_{mo}}{\rho_{po}}(1 - e^{-\frac{k}{r_{bi}^a}}) \tag{10}$$

where $V_p$ is the volume of polymer, $\rho_{po}$ is the density of polymer (1.18 g/ml for pBzMA, 1.15 g/ml for pHEMA), $\rho_{mo}$ is the density of monomer (1.04 g/ml for BzMA, 1.07 g/ml for HEMA), $k$ is the composite kinetic parameter ($k = k_C$ in the collision model, and $k = k_D$ in the diffusion model) and $a$ represents the scaling of the base radius ($a = 1/2$ in the collision model, and $a = 1$ in the diffusion model). Our experimental results plotted in Fig. 4B,C fit well to the scaling laws between the polymer volume fraction and the base radius based on equation (10), indicating both Knudsen and Fickian diffusion likely happen simultaneously in CDP as described above.

Using the expression derived from the kinetics of polymerization and primary radical diffusion, we can finally predict the height and base radius of the polymer microdomes based on the monomer droplet base radius and synthesis conditions. By combining equation (3), (4) and (10), we obtain:

$$(H_f)_{model} = (1.18*\lambda)^{\frac{1}{2}}\left(\frac{\rho_{mo}}{\rho_{po}}\left(1 - e^{-\frac{k}{r_{bi}^a}}\right)\right)^{1/2} r_{bi} = 1.02\lambda^{\frac{1}{2}}\left(1 - e^{-\frac{k}{r_{bi}^a}}\right)^{1/2} r_{bi} \tag{11}$$

$$\left(r_{bf}\right)_{model} = \left(\frac{1.18}{\lambda}\right)^{\frac{1}{4}} \left(\frac{\rho_{mo}}{\rho_{po}}\left(1 - e^{-\frac{k}{r_{bi}^a}}\right)\right)^{1/4} r_{bi} = 1.01\lambda^{-\frac{1}{4}}\left(1 - e^{-\frac{k}{r_{bi}^a}}\right)^{1/4} r_{bi} \quad (12)$$

where the kinetic factors $k$ and $a$ can be determined from the fitting and model choice in Fig. 4B,C, and the geometric factor, $\lambda$, can be derived from the model fitting shown in Fig. 3E. By incorporating these factors, we find excellent agreement between the experimentally measured dome height $(H_f)_{exp}$ and base radius $(r_{bf})_{exp}$ with the model-predicted height $(H_f)_{model}$ and radius $(r_{bf})_{model}$ from equations (11) and (12), respectively (Fig. 4D,E; Table S1).

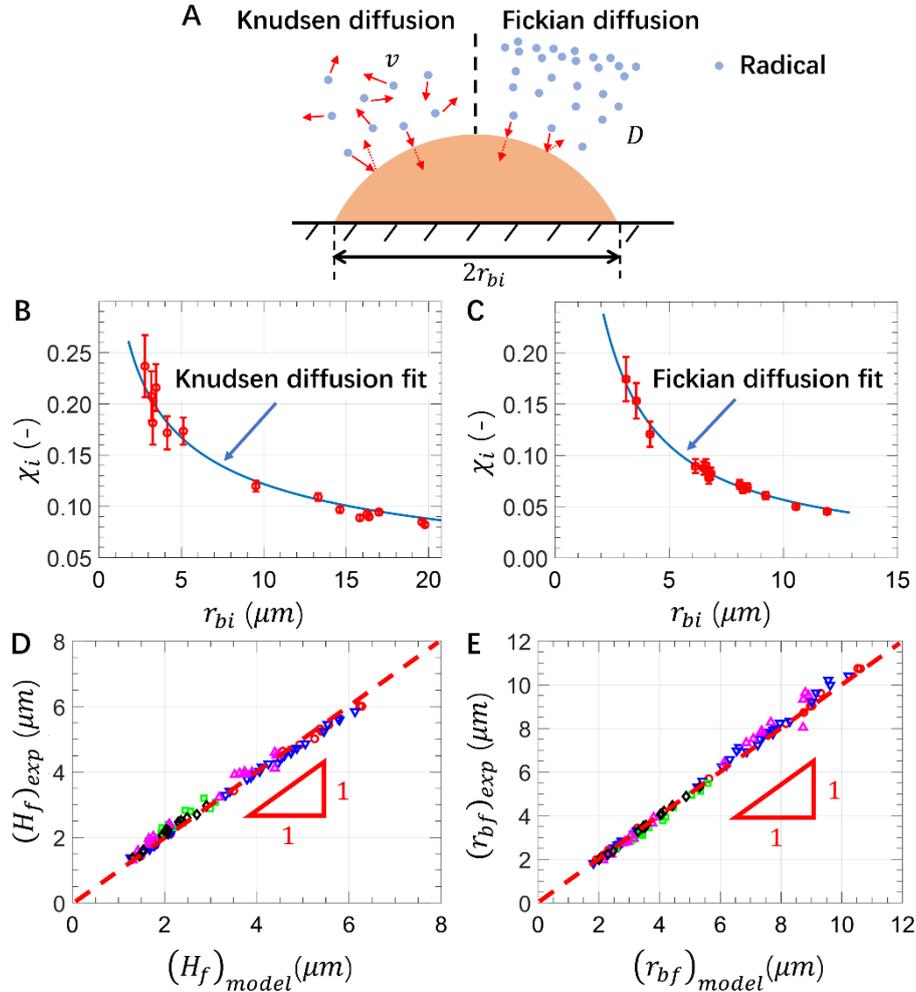

**Fig. 4 | Model development and validation for the kinetic processes of primary radical diffusion and polymerization, which enables prediction of the microdome geometric parameters.** (**A**) Schematic of the primary radical diffusion during the polymerization process in CDP. The radicals diffuse in the vapor phase via two possible mechanisms: Knudsen diffusion and Fickian diffusion, which can be characterized by the average velocity of radicals $v$ and the diffusion coefficient $D$, respectively. (**B,C**) Plot of the polymer volume fraction $(\chi_i)$ versus droplet base radius $(r_{bi})$ at the onset of evaporation. One fits the Knudsen diffusion model, while the other one fits the Fickian diffusion model. (**D,E**) The dome height $(H_f)_{exp}$ and dome base radius $(r_{bf})_{exp}$ agree with the model prediction $(H_f)_{model}$ and $(r_{bf})_{model}$ that are obtained by the geometric relation and the model of CDP kinetics, irrespective of the initial droplet radius, monomer species, and polymerization time, which are represented with different data symbols in the figure (see details in Supplementary Note 6).

**Biphilic microdome arrays deter biofilm formation by *P. aeruginosa***

Microdome-shaped morphological surfaces are commonly observed in marine organisms. For instance, the crustacean *Cancer pagurus* exhibits a dome-like surface structure with microscale texture, which has been postulated to resist the settlement of foulers like diatoms(*58*). Inspired by such morphological features, we created biphilic microdome arrays using HEMA as the monomer.

To evaluate the antifouling performance of the biphilic microdome arrays, we conducted a fouling experiment using *P. aeruginosa* strain PAO1 as the model organism for its abundant Type IV pili (TFP) appendages (responsible for initiating surface attachment), and its known ability to readily produce biofilms(*27*). Uncoated, pHEMA-coated, and pPFDA-coated glass coverslips and the ones bearing pHEMA microdome arrays on a pPFDA background were incubated in freshly diluted (1:100) PAO1 cultures for 8 h to allow bacterial adhesion and growth, followed by thorough rinsing to remove any loosely attached bacteria. At the end of the incubation, the biofilms were subjected to two characterization methods: (i) staining with crystal violet and quantification of the biomass based on absorbance at 550 nm (Fig. 5B), and (ii) staining with SYTO9 and confocal microscopic imaging (Fig. 5C).

The biofilm growth on the hydrophilic surfaces, i.e., uncoated and pHEMA-coated glass slides, is significantly lower than that on the hydrophobic surface, i.e., pPFDA coating. This is consistent with previous reports that credit the enthalpic penalty associated with the displacement of surface-bonded water molecules by biofoulers for the effective biofouling deterrence of hydrophilic surfaces(*26, 59*). Interestingly, the biphilic microdome arrays led to reduced biofilm formation to a degree similar to that on pHEMA-coated surfaces (Fig. 5B). This antifouling performance is surprising given the high water contact angle on the pHEMA microdome arrays, i.e., 105.4±0.7° (Fig. 5A), which is close to the water contact angle on the pPFDA coating (115.8±0.5°, Fig. 5A). In contrast, pHEMA thin films exhibit a water contact angle of 13.3±0.2°, much lower than that on the pHEMA microdome arrays despite their comparable antifouling performance. Additionally, the receding angle of the pHEMA microdome arrays is 40.3±0.6°, which is also higher than pHEMA thin film, i.e., 11.9±0.1° (see detailed results in Supplementary Note S7).

The biphilic microdome arrays point to a fresh antifouling mechanism, one that hints at the effectiveness of discrete biphilic morphological features to deter biofilm formation and that a continuous hydration layer may not be a prerequisite for fouling deterrence. To further understand the antifouling mechanism of the biphilic microdome arrays, we include SEM images showing the sub-micro domes (Fig. 5D), which are analyzed using the quantitative model described in the preceding sections. As shown in Fig. 5E, the dome base radius ($r_{bf}$) in the sub-micro range (300 nm to 900 nm) aligns well with the model prediction based on equation (12). These domes, with sizes comparable to an individual bacterial cell, introduce discontinuities in the surface energy landscape and reduce the area conducive to bacterial adhesion. While morphology-enabled biofouling deterrence has been reported(*60, 61*), those reports invariably focus on uniformly hydrophilic or hydrophobic surfaces (for hydration-based adhesion mitigation or self-cleaning properties stemming from the Cassie-Baxter wetting state), hence distinct from the biphilic microdome arrays reported here.

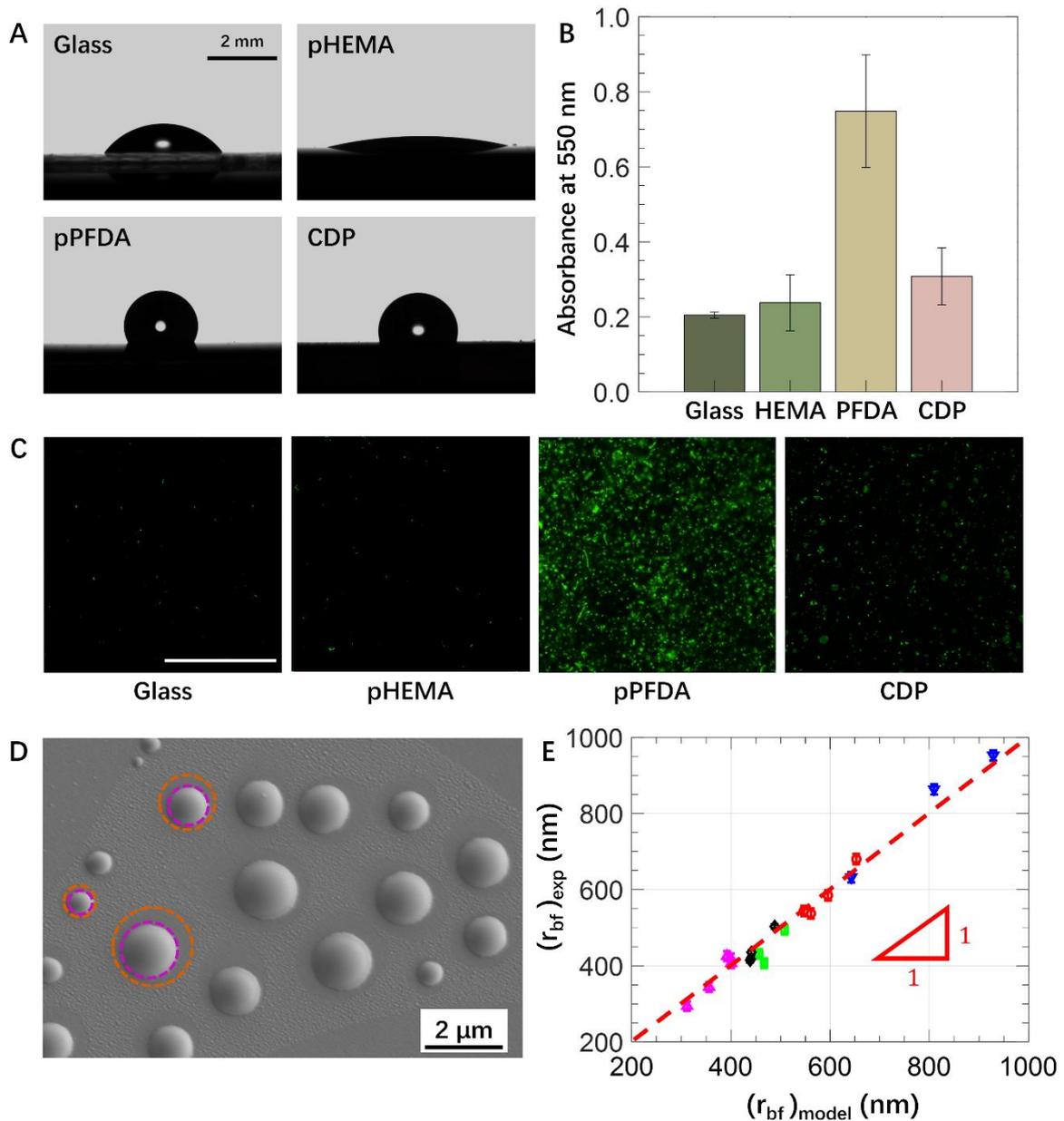

**Fig. 5 | Antifouling performance of the biphilic microdome array surfaces.** (**A**) Water contact angle measurement on uncoated, pHEMA-coated, pPFDA-coated, and pHEMA microdome-array covered glass slides. The contact angles are 52.8±0.7°, 13.3±0.2°, 115.8±0.5° and 105.4±0.7°, respectively. (**B**) Absorbance (at 550 nm) of *P. aeruginosa* biofilms (stained using crystal violet) after 8 h of incubation on the four aforementioned surfaces. (**C**) Representative fluorescent micrographs (out of 5 images, see Supplementary Fig. S4 for other fluorescent micrographs) of the *P. aeruginosa* biofilms after 8 h of incubation with the four aforementioned surfaces. The scale bar is 100 μm. (**D**) SEM images of sub-micro arrays. The orange dashed circles indicate the initial droplet base, while the purple dashed circles denote the polymer dome base. (**E**) The dome base radius $(r_{bf})_{exp}$ captured experimentally agrees with the model prediction $(r_{bf})_{model}$ for the biphilic sub-micro dome array (300 nm to 900 nm).

## Discussion

In summary, we show that despite the highly dynamic nature of condensation and evaporation, CDP can be leveraged to synthesize biphilic microdome arrays with predictable geometric parameters, enabled by the physics-based model. The model connects the diameter and height of each microdome to the synthesis conditions and an observable and controllable quantity, i.e., the radius of the microdroplet. It was derived based on the geometric relation and scaling laws, while integrating polymerization kinetics and polymer dome shape analysis. The size and shape of the polymer dome are strongly dependent on the polymer volume fraction in the droplet ($\chi_i$), with the dome base radius ($r_{bf}$) scaling as $r_{bf} \sim \chi_i^{1/4}$, and the dome height scaling as $H_f \sim \chi_i^{1/2}$. The polymer volume fraction is in turn predictable based on the droplet size ($r_{bi}$), following the scaling law of $\chi_i \sim \left(1 - e^{-kr_{bi}^{-a}}\right)$, where $a$=1/2 under Knudsen diffusion, and $a$=1 under Fickian diffusion. Additionally, we fabricated biphilic microdome arrays comprising pHEMA microdomes on a pPFDA thin film, which reduces biofilm formation by *P. aeruginosa* despite its overall hydrophobicity, pointing to a fresh design rationale for antifouling surfaces. This work exemplifies a paradigm where physics-based modeling can be leveraged to improve the precision of chemical synthesis. In this example, it enables the precision synthesis of biphilic microdome arrays in a template-free fashion, which has challenged existing synthesis and fabrication methods. This synthesis capability, in turn, points to impactful applications in water harvesting, heat transfer, bioadhesion control, drug delivery, and injectable implants.

## Materials and methods

### Materials
2-hydroxyethyl methacrylate (HEMA, >99%), benzyl methacrylate (BzMA, 96%), *tert*-butyl peroxide (TBPO, >98%), *1H,1H,2H,2H*-perfluorodecyl acrylate (PFDA, >97%) were purchased from Sigma-Aldrich. All chemicals were used as-is without further purification.

### Polymer thin film synthesis
Silicon wafers were coated with a thin film (50–100 nm) using the iCVD technique. Briefly, silicon wafers were placed on a stage maintained at 30°C in the iCVD reactor chamber. PFDA, heated to 80°C, was introduced into the reactor at a flow rate of 0.15 sccm, while TBPO was flowed in at 0.6 sccm. During deposition, the chamber pressure was maintained at 400 mTorr, with the filament array heated to 300°C. Film growth on the Si substrate was monitored by in situ interferometry, using a HeNe laser source (wavelength = 633 nm; JDS Uniphase).

Glass slides (VWR 48368-040, 18 mm × 18 mm, No. 2) were coated with a thin film (50–100 nm) using the same iCVD technique as for silicon wafers described above. PFDA, heated to 80°C, was introduced into the reactor using the following conditions: PFDA, TBPO flow rates of 0.15, 0.6 sccm, respectively. The chamber pressure was maintained at 400 mTorr during the deposition. HEMA, heated to 80°C, was introduced into the reactor using the following conditions: HEMA, TBPO flow rates of 0.2, 0.3 sccm, respectively. The chamber pressure was maintained at 300 mTorr during the deposition.

### Condensed droplet polymerization procedure
CDP was carried out in a cylindrical vacuum reactor equipped with a thermoelectric cooler (TEC) module. Prior to initiating CDP, a pPFDA-coated silicon wafer/glass slide was placed on the TEC, which served to locally control the substrate temperature. The reactor was evacuated to a base pressure of ≤5 mTorr to establish high-vacuum conditions. The CDP process consisted of three sequential steps: condensation, polymerization, and evaporation. In the condensation step, the vacuum line was closed, and monomer vapor (BzMA/HEMA) was introduced into the chamber at a flow rate of 0.2 sccm. The TEC was activated to cool the substrate, creating a localized temperature gradient that promoted condensation of monomer vapor into microdroplets on the substrate surface. Condensation was allowed to proceed until the droplets reached the desired size, at which point monomer flow was terminated to stop this step. Next, in the polymerization step, tert-butyl peroxide (TBPO) was introduced at a flow rate of 0.6 sccm. As the TBPO vapor passed through a heated zone maintained at ~300 °C, it thermally decomposed into *tert*-butoxyl and methyl radicals. These gas-phase radicals initiated polymerization within the condensed droplets. Polymerization proceeded for a fixed duration (1-4 min), after which the TBPO flow was stopped. In the final evaporation step, vacuum was reapplied and the substrate temperature was gradually increased. This removed the unreacted monomer from the monomer–polymer mixture droplets, resulting in the formation of solid polymer microdomes on the substrate.

**Material characterization**
FTIR spectroscopy was conducted using a Bruker VERTEX Series V80v spectrometer operating in transmission mode. Spectra were collected in the range of 400–4000 $cm^{-1}$ at a spectral resolution of 4 $cm^{-1}$, using a mercury cadmium telluride detector. An uncoated silicon wafer was used as the background reference. Each spectrum represents an average of 128 scans and was baseline-corrected prior to analysis.

SEM and EDX analyses were performed using a Zeiss Gemini SEM 500 operated at an accelerating voltage of 3–4 kV. Prior to SEM imaging, samples were sputter-coated with approximately 3 nm of gold/palladium to enhance surface conductivity. For EDX measurements, samples were sputter-coated with approximately 3 nm of carbon.

Contact angle measurements and droplet profile imaging were performed using a Rame-Hart Model 500 contact angle goniometer. For static contact angle measurements, 5 µL droplets of deionized water, HEMA and BzMA were dispensed onto the substrate using a micropipette. Dynamic contact angle measurements were conducted by gradually dispensing the droplet up to 10 µL and then retracting it, while recording contact angle values at 0.5 s intervals.

Atomic force microscopy (AFM) was performed using an Asylum Research MFP-3D-BIO system to characterize the morphology of the polymer microdomes. Scans were conducted in AC tapping mode across 20 × 20 µm areas at a scan rate of 0.5 Hz. The profiles of the microdomes, with diameters in the range of 2−8 µm, were traced across the center of the dome to determine the end-to-end distance and base-to-tip height.

**Dome height measurement**
The CDP samples were coated using a Varian-Thermal-Bell-Jar Evaporator with a thin film (~20 nm) of gold at a deposition rate of 1 Å/s, making the sample surfaces highly reflective. Keyence VK-X260 Laser-Scanning Profilometer was used to capture the dome height profile. We used 150X magnification with planar and vertical resolutions of 0.13 µm and 0.08 µm, respectively.

**Antifouling performances evaluation**

The antifouling performance of the biphilic surfaces was evaluated using *P. aeruginosa* (PAO1 strain), a commonly used model organism in biofilm and surface fouling studies. The strain was preserved in 50% glycerol at -80°C prior to use. For the experiment, the bacteria were streaked onto fresh LB agar plates, and incubated overnight (around 16 hours) at 37°C. A single colony from the overnight culture was inoculated in LB medium and cultured for 18 hours at 37°C, 225 rpm, reaching an $OD_{600}$ of ~0.5-0.6. This culture was then diluted 100 times in fresh LB medium for use in the antifouling experiment. Bare glass coverslips (VWR 48368-040, 18 mm × 18 mm, No. 2) and coated samples were placed into 6-well plates (Corning, Costar® 3516) and sterilized under UV light for 1 hour on each side. 3 ml of the diluted bacterial suspension was added to each well, ensuring full immersion of the coverslips. The plates were incubated for 8 hours to ensure substantial biofilm formation. After the incubation, the coverslips were gently rinsed three times with a 0.15M NaCl solution to remove loosely attached bacteria. The biofilm on the surfaces was assessed by crystal violet staining and confocal fluorescence microscopy imaging. For crystal violet staining, the rinsed fouled coverslips were submerged in 0.5% (*w/v*) crystal violet solution for 15 min, washed three times with Milli-Q water and air-dried overnight. The coverslips were then submerged in 2 ml 33% (*v/v*) acetic acid solution in a 6-well plate to dissolve the stained biofilm. 150 μL of each solution was transferred to a 96-well plate for absorbance measurement at 550 nm, along with two control samples of acetic acid solution for background correction. For confocal microscopy imaging, the fouled coverslips were stained using a STYO9-containing LIVE/DEAD BacLight Bacterial Viability Kit (Invitrogen, L7007A) for 15 min, rinsed with NaCl solution, and then put onto a confocal dish for subsequent imaging using a Zeiss i880 confocal microscope (40× oil objective).

**Acknowledgments**

This material is based upon work supported by the National Science Foundation (NSF) through the Faculty Early Career Development Program (CAREER) under Grant No. CMMI-2144171. It is also supported by the Department of the Navy, Office of Naval Research under ONR award N00014-23-1-2189 and the Camille Dreyfus Teacher-Scholar Award from the Camille and Henry Dreyfus Foundation.

**Author contributions**

R.Y. conceived the idea and supervised the work. H.X. conducted CDP syntheses, analyzed, and interpreted the data. H.X. developed the model. H.X. and H.S. performed the fouling experiment. H.X. and R.Y. drafted the manuscript and edited it for critically important intellectual content.

# Supplementary information of "Template-free precision synthesis of biphilic microdome arrays"

## Content

## S1. Evaporation mode analysis

From our observation on the squared average droplet contact radius versus time plot ($r_b^2$ vs $t$ in Fig. 2d), we captured a linear behavior in stages I and II, suggesting that evaporation likely occurs in a constant contact angle mode by doing the following analysis. For a droplet evaporating in constant contact angle mode, the temporal base radius can be expressed by equation (1): $r_b^2 = r_{bi}^2 - \frac{4D(c_S - c_\infty)\sin^2\theta}{\rho_L(1-\cos\theta)(2+\cos\theta)}t$, where $r_{bi}$ is the initial droplet base radius, $D$ is the self-diffusion coefficient of the monomer vapor in the chamber, $c_S$ is the vapor concentration at the spherical cap of the droplet, $c_\infty$ is the vapor concentration in the chamber, $\theta$ is the droplet contact angle, $\rho_L$ is the density of the monomer. During the evaporation process, the vapor diffusion coefficient $D$ and monomer density $\rho_L$ didn't change. For the vapor concentration terms, we can calculate them by $c = PM_w/RT$ assuming ideal gas behavior. For the vapor concentration in the chamber, we use the monomer pressure in the chamber (P ~ 5 mTorr) to do the calculation, $c_\infty = 5.0 \times 10^{-5}$ g/cm³. While for the vapor concentration at the spherical cap of the droplet $c_S$, we use monomer vapor pressure and the temperature of TEC to do the calculation. Vapor pressure $P_v$ is a function of the temperature of droplet, and can be described by the Clausius-Clapeyron Equation: $\ln P = A/T + B$. For BzMA, $A = -7338.15$, $B = 26.13$. Since the droplet size is very small compared to the TEC, we can assume that the droplet temperature equals the TEC temperature. In stage I, the TEC temperature was kept at 10 °C, consequently, $P_v = 9.2$ mTorr, $c_S = 9.5 \times 10^{-5}$ g/cm³. In stage II, the TEC temperature was adjusted and kept at 20 °C, with a corresponding vapor pressure $P_v = 22.1$ mTorr, vapor concentration $c_S = 2.14 \times 10^{-5}$ g/cm³. Thus, we obtain that $4D(c_S - c_\infty)/\rho_L$ are unchanged in both stages I and II, and to achieve linear behavior of $r_b^2$ vs $t$, the term, $\sin^2\theta/(1-\cos\theta)(2+\cos\theta)$, should also be unchanged, which indicates the contact angle is constant in stage I and II.

## S2. Dome height measurement comparison

To measure the height of polymer domes, we use two different methods: atomic force microscopy (AFM) shown in the Fig. S1a and Keyence VK-X260 Laser-Scanning Profilometer (Profilometer) shown in Fig. 2b. The advantage of using Profilometer is its scanning area, which enables the high-throughput measurement of the height of a large number of domes by doing a single measurement. Since the polymer dome size is in microns, AFM measurement becomes time-consuming and low throughput. By collecting height and radius data through these two measurements, we found their results agree well with each other (Fig. S1b). Thus, we use Profilometer to effectively measure dome height and radius.

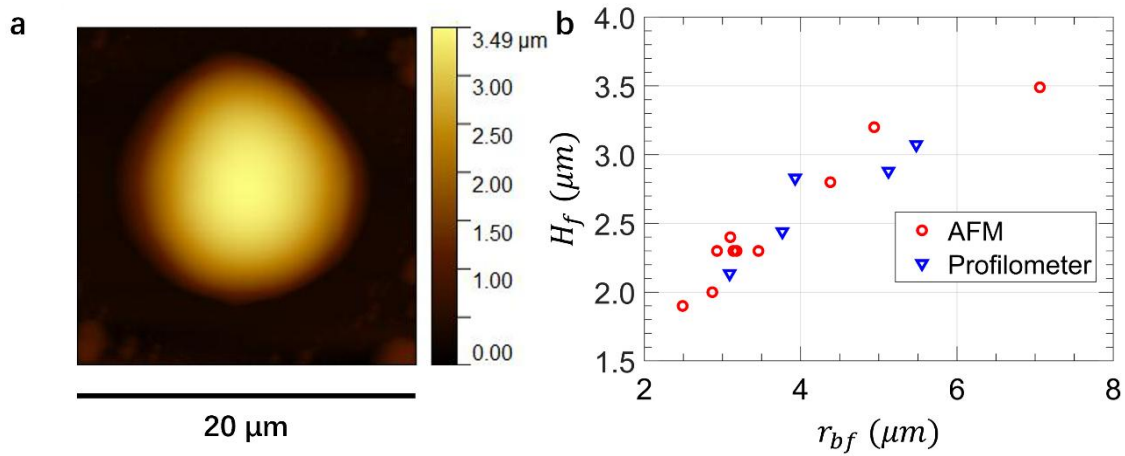

**Fig. S1 | Dome height measurement and comparison. a,** Height of a polymer dome measured by atomic force

microscopy (AFM). **b,** Comparison of the dome height measurements. Red circles represent results from AFM and blue triangles represent results from the Keyence VK-X260 Laser-Scanning Profilometer (Profilometer). These two sets of data show the same trend, proving that the measurement of profilometer is reliable.

## S3. Dome shape and model validation

Based on direct measurements of the dome base radius $r_{bf}$ and height $H_f$, we examined the dome geometry in nondimensionalized space. We found that the dimensionless height scales quadratically with dimensionless base radius $H_f/r_{bi} = \lambda(r_{bf}/r_{bi})^2$, with a fitted coefficient $\lambda = 0.75$ as shown in Fig. 3e, suggesting the dome shape can be approximated by a parabola. To further validate this geometric assumption, we plotted $(H_f/r_{bi})^2$ versus $1.18\chi_i$ based on our model described in equation (3), $(H_f/r_{bi})^2 = \lambda(1.18\chi_i)$. As shown in Fig. S2, the slope of the linear fit equals 0.78, which is very close to the 0.75 previously obtained, suggesting the dome shape approximation and our model are reasonable.

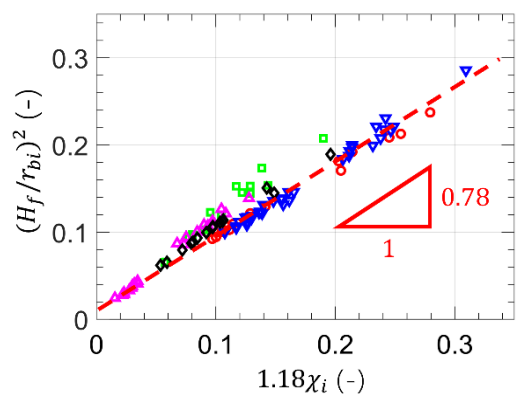

**Fig. S2 | Dome shape validation.** Dimensionless dome height square $(H_f/r_{bi})^2$ scales with polymer volume fraction times 1.18 ($1.18\chi_i$) with a slope of 0.78, irrespective of initial droplet radius, monomer species and polymerization time, which are represented with different data symbols in the figure (see Section S6 for the detailed synthesis conditions corresponding to each symbol).

## S4. Contact angle measurement

A 5 µl monomer droplet was deposited on a pPFDA-coated silicon wafer. Contact angle measurements were performed using a Rame-Hart Model 500 contact angle goniometer including the image capture and contact angle analysis. The contact angle for BzMA and HEMA are 86.1°±0.1° and 87.4°±0.1° respectively as shown in Fig. S3a,b.

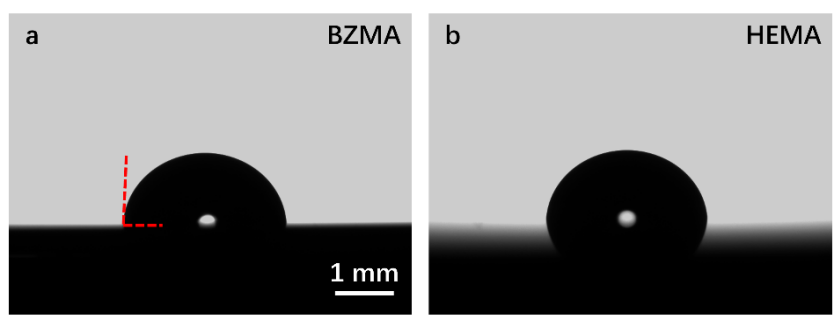

**Fig. S3 | Contact angle measurement. a,** Contact angle of BzMA on pPFDA is 86.1°±0.1°. **b,** Contact angle of HEMA on pPFDA is 87.4°±0.1°.

## S5. Theoretical calculation for kinetics modeling during polymerization

In the main text, we derived equation (6) describing the rate of polymerization. $R_p = -\frac{d[M]}{dt} = k_p[M]\left(\frac{R_i}{2k_t}\right)^{\frac{1}{2}}$, where $R_i = \frac{1}{r_b}[R\bullet]_v v \Gamma \frac{\alpha_i}{\beta_i}$ can be described by equation (7) in the main text (under the collision model). $[R\bullet]_v$ is the radical concentration in the vapor phase and can be calculated by integrating both sides of $d[R\bullet]_v = 2fk_d[I]dt$. Since the initiator is introduced in constant flow rate (0.6 sccm), we define $[I] = Qt$, where $Q$ is the initiator flow rate. Thus, we obtain an expression for $[R\bullet]_v$ as a function of time:

$$[R\bullet]_v = \int_0^t 2fk_d Qt dt = fk_d Qt^2 \qquad (13)$$

where $f$ is the initiator efficiency, $k_d$ is the rate constant of TBPO decomposition. We then apply equation (S1) to rate of initiation $R_i$:

$$R_i = \frac{1}{r_{bi}}[R\bullet]_v v \frac{\alpha_i}{\beta_i} = \frac{\alpha_i}{\beta_i}\frac{1}{r_{bi}} v \Gamma fk_d Qt^2 \qquad (14)$$

where $\alpha_i/\beta_i$ is the geometrical factor, $r_{bi}$ is the initial base radius of droplet, $v$ is the average velocity of radicals under Knudsen diffusion. Then, we can apply the expression in equation (S2) to equation (6) and integrate both sides to calculate the unreacted monomer concentration after polymerization $[M]_{t_{po}}$:

$$\int_{[M]_0}^{[M]_{t_{po}}} -\frac{1}{[M]}d[M] = \int_0^{t_{po}} k_p \left(\frac{1}{2k_t}\frac{\alpha_i}{\beta_i}\frac{1}{r_b}v\Gamma fk_d Q\right)^{\frac{1}{2}} t dt$$

$$[M]_{t_{po}} = [M]_0 e^{-\frac{1}{2}k_p\left(\frac{1}{2k_t}\frac{\alpha_i}{\beta_i}v\Gamma fk_d Q\right)^{\frac{1}{2}}t_{po}^2 \frac{1}{r_{bi}^{1/2}}} = [M]_0 e^{-\frac{k_C}{r_{bi}^{1/2}}} \qquad (15)$$

where $t_{po}$ is the time for polymerization, $[M]_0$ is the initial monomer concentration in droplet, $k_C = \frac{1}{2}k_p\left(\frac{1}{2k_t}\frac{\alpha_i}{\beta_i}v\Gamma fk_d Q\right)^{\frac{1}{2}}t_{po}^2$ is the kinetics factor under Knudsen diffusion (i.e., collision model). The physical quantity in $k_C$ only depends on the reactor conditions and chemical properties of monomer and initiator, and it remains the same for all droplets since our sample is very small (< 1 cm×1 cm) compared to the reactor. Thus, the initial droplet size is the only variable affecting the dome size and shape under identical reaction condition and using the same reaction system.

The derivation above addresses the case of Knudsen diffusion, whereas in the case of Fickian diffusion (i.e., the diffusion model), the only difference is the expression for the radical flux, $\Phi = D[R\bullet]_v/r_{bi}$, where $D$ is the diffusion coefficient of radicals in reactor and $[R\bullet]_v$ is derived in equation (S1). The rate of initiation in monomer droplet can be derived consequently:

$$R_i = -\frac{d[R\bullet]}{dt} = \frac{1}{V_i}\frac{1}{N_A}\Phi A = \frac{1}{r_{bi}^2}[R\bullet]_v D\Gamma \frac{\alpha_i}{\beta_i} = \frac{\alpha_i}{\beta_i}\frac{1}{r_{bi}^2}D\Gamma fk_d Qt^2 \qquad (16)$$

Then, we can follow the same derivation process as we did in the collision model, thus, we get the

unreacted monomer concentration after polymerization for Fickian diffusion:

$$[M]_{t_{po}} = [M]_0 e^{-\frac{1}{2}k_p\left(\frac{1}{2k_t}\frac{\alpha_i}{\beta_i}D\Gamma f k_d Q\right)^{\frac{1}{2}} t_{po}^2 \frac{1}{r_{bi}}} = [M]_0 e^{-\frac{k_D}{r_{bi}}} \tag{17}$$

where $k_D = \frac{1}{2}k_p\left(\frac{1}{2k_t}\frac{\alpha_i}{\beta_i}D\Gamma f k_d Q\right)^{\frac{1}{2}} t_{po}^2$ is the kinetics factor for the diffusion model.

## S6. Synthesis conditions for the CDP microdomes

Microdome synthesis was conducted under varying conditions, including differences in initial droplet radius, monomer species, and polymerization time. Distinct marker styles were used to label each condition in Fig. 3e–g and Fig. 4d,e. The corresponding experiment conditions are summarized in Table S1.

**Table S1. Synthesis conditions for CDP domes.** Microdome synthesis was performed under a range of conditions, including variations in initial droplet radius, monomer species and polymerization time. Different marker styles used in Fig. 3 and Fig. 4 correspond to respective synthesis conditions.

| Marker style | Initial droplet radius | Monomer species | Polymerization time |
|---|---|---|---|
|  | 2.8-19.8 μm | BzMA | 2 min |
|  | 2.6-18.6 μm | BzMA | 4 min |
|  | 4.2-12.0 μm | BzMA | 4 min |
|  | 3.7-27.1 μm | BzMA | 2 min |
|  | 3.2-11.9 μm | HEMA | 1 min |

## S7. Receding contact angle measurement

Receding water contact angle measurements were performed using a Rame-Hart Model 500 contact angle goniometer including the image capture and contact angle analysis. The measurements for receding angle of the glass slides, pHEMA-coated glass slides, pPFDA-coated glass slides and biphilic microdome array surfaces are summarized in Table S2.

**Table S2. Receding water contact angles.** The receding angle of water on glass slides, pHEMA-coated glass slides, pPFDA-coated glass slides and biphilic microdome array surfaces.

| Sample | Receding angle (°) |
|---|---|
| Glass slides | 26.7±0.3 |
| pHEMA-coated glass slides | 11.9±0.1 |
| pPFDA-coated glass slides | 82.2±0.3 |
| Biphilic microdome array surfaces | 40.3±0.6 |

## S8. Confocal microscope images for samples underwent biofouling experiments

Five fluorescent micrographs were acquired by confocal microscopy for each of the four aforementioned surfaces. One representative image per surface is presented in Fig. 5c, with the remaining micrographs provided in Fig. S4.

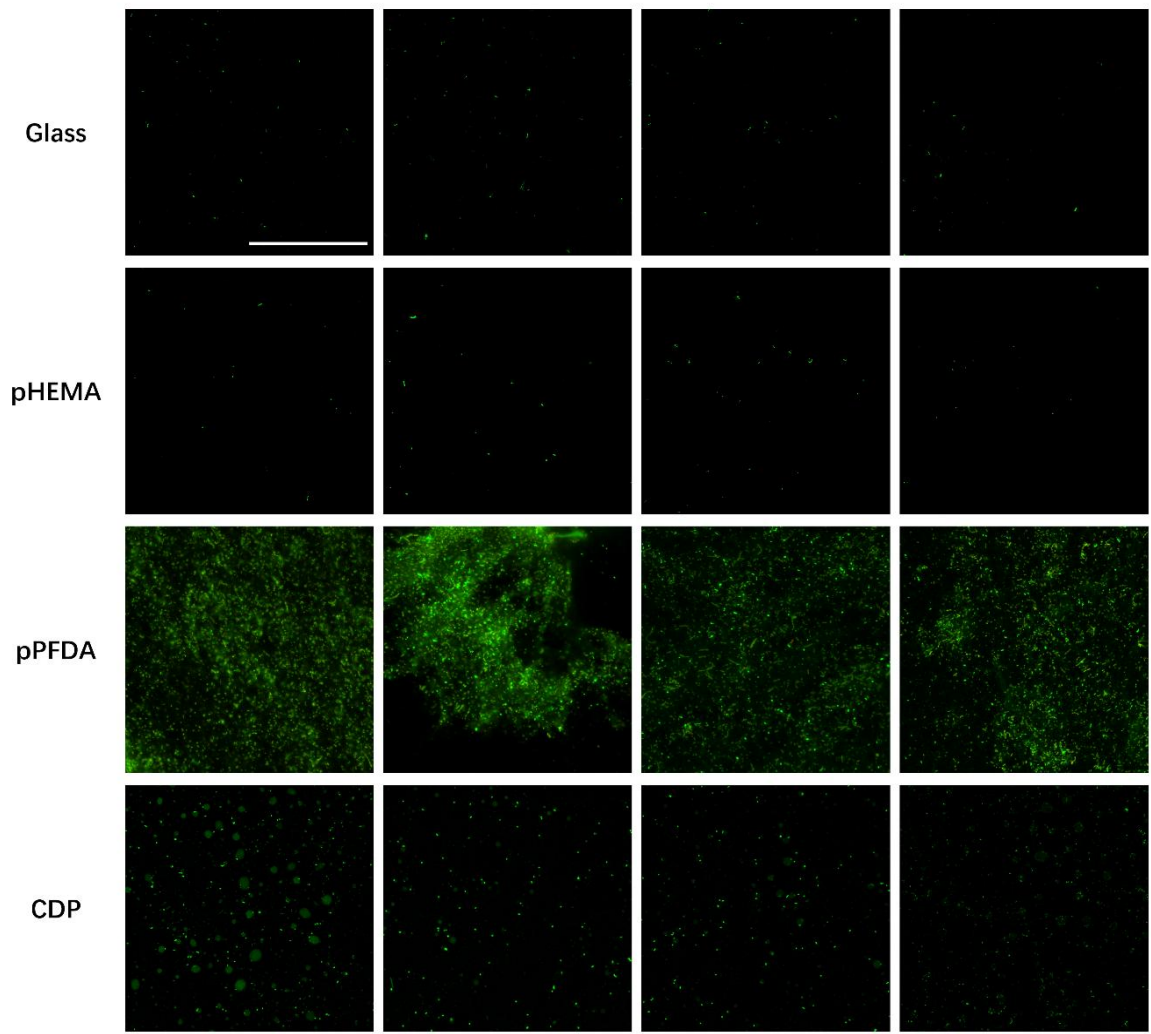

**Fig. S4 | Confocal fluorescence microscopy images of the four aforementioned surfaces.** For each surface, four images were captured using confocal microscopy. The scale bar is 100 μm.